\documentclass[12pt]{article}
\usepackage[dvips]{graphicx}

\renewcommand{\bar}[1]{\overline{#1}}

\newcommand{\ket}[1]{\,\left|\,{#1}\right\rangle}

\begin {document}
\begin{flushright}
{\small
SLAC--PUB--11976\\
July 2006\\}
\end{flushright}

\begin{center}
{{\bf\LARGE NEW PERSPECTIVES FOR QCD\\[1ex] FROM AdS/CFT}\footnote{Work
supported by Department of Energy contract DE--AC02--76SF00515.}}

\bigskip
\bigskip
{\it Stanley J. Brodsky \\
Stanford Linear Accelerator Center \\
Stanford University, Stanford, California 94309 \\
E-mail:  sjbth@slac.stanford.edu}
\medskip
\end{center}

\vfill
\begin{center}
{\bf\large Abstract }
\end{center}

The AdS/CFT correspondence between conformal field theory  and
string states in an extended space-time  has provided new insights
into not only  hadron spectra, but also their light-front
wavefunctions. We show that there is an exact correspondence between
the fifth-dimensional coordinate of anti-de Sitter space $z$ and a
specific impact variable $\zeta$ which measures the separation of
the constituents within the hadron in ordinary space-time. This
connection allows one to predict the form of the  light-front
wavefunctions of mesons and baryons, the fundamental entities which
encode hadron properties and scattering amplitudes. A new
relativistic Schr\"odinger light-cone equation is found which
reproduces the results obtained using the fifth-dimensional theory.
  \vfill

\begin{center}
{\it Presented at the \\ Workshop on
Continuous Advances in QCD\\
11--14 May 2006  \\
Minneapolis, Minnesota, USA  }
\end{center}

\vfill \newpage

\section{The Conformal Approximation to QCD}

One of the most interesting recent developments in hadron physics
has been the use of  Anti-de Sitter space holographic methods in
order to obtain a first approximation to nonperturbative QCD. The
essential principle underlying the AdS/CFT approach to conformal
gauge theories is the isomorphism of the group of Poincare' and
conformal transformations $SO(4,2)$ to the group of isometries of
Anti-de Sitter space.  The AdS metric is
\[ds^2 = {R^2\over z^2}(\eta^{\mu \nu} dx_\mu
dx^\mu - dz^2)\]
which is invariant under scale changes of the
coordinate in the fifth dimension $z \to \lambda z$ and $ dx_\mu \to
\lambda dx_\mu$.  Thus one can match scale transformations of the
theory in $3+1$ physical space-time to scale transformations in the
fifth dimension $z.$ The amplitude $\phi(z)$ represents the
extension of the hadron into the fifth dimension.  The behavior of
$\phi(z) \to z^\Delta$ at $z \to 0$ must match the twist dimension
of the hadron at short distances $x^2 \to 0.$   As shown by
Polchinski and Strassler\cite{Polchinski:2001tt}, one can simulate
confinement by imposing the condition $\phi(z = z_0 ={1\over
\Lambda_{QCD}}).$  This approach, has been successful in reproducing
general properties of scattering processes of QCD bound
states\cite{Polchinski:2001tt,Brodsky:2003px}, the low-lying hadron
spectra\cite{deTeramond:2005su,Erlich:2005qh}, hadron couplings and
chiral symmetry breaking\cite{Erlich:2005qh,Hong:2005np}, quark
potentials in confining backgrounds\cite{Boschi-Filho:2005mw} and
pomeron physics\cite{Boschi-Filho:2005yh}.

It was originally believed that the AdS/CFT mathematical tool could
only be applicable to strictly conformal theories such as $\mathcal{N}=4$
supersymmetry.  However, if one considers a semi-classical
approximation to QCD with massless quarks and without particle
creation or absorption, then the resulting $\beta$ function is zero,
the coupling is constant, and the approximate theory is scale and
conformal invariant. Conformal symmetry is of course broken in
physical QCD; nevertheless, one can use conformal symmetry as a {\it
template}, systematically correcting for its nonzero $\beta$
function as well as higher-twist effects. For example,
``commensurate scale relations"\cite{Brodsky:1994eh}
 which relate QCD observables to each
other, such as the generalized Crewther relation\cite{Brodsky:1995tb},
 have no
renormalization scale or scheme ambiguity and retain a convergent
perturbative structure which reflects the underlying conformal
symmetry of the classical theory.  In general, the scale is set such
that one has the correct analytic behavior at the heavy particle
thresholds\cite{Brodsky:1982gc}.

In a confining theory where gluons have an effective mass, all
vacuum polarization corrections to the gluon self-energy decouple at
long wavelength.  Theoretical\cite{Alkofer:2004it} and
phenomenological\cite{Brodsky:2002nb} evidence is in fact
accumulating that QCD couplings based on physical observables such
as $\tau$ decay\cite{Brodsky:1998ua} become constant at small
virtuality;  {\em i.e.}, effective charges develop an infrared fixed
point in contradiction to the usual assumption of singular growth in
the infrared. The near-constant behavior of effective couplings also
suggests that QCD can be approximated as a conformal theory even at
relatively small momentum transfer. The importance of using an
analytic effective charge\cite{Brodsky:1998mf} such as the pinch
scheme\cite{Binger:2006sj,Cornwall:1989gv} for unifying the
electroweak and strong couplings and forces is also
important\cite{Binger:2003by}.  Thus conformal symmetry is a useful
first approximant even for physical QCD.

\section{Hadronic Spectra in AdS/QCD}

Guy de Teramond and I\cite{Brodsky:2006uq,deTeramond:2005su} have
recently shown how a holographic model based on truncated AdS space
can be used to obtain the hadronic spectrum of light quark $q \bar
q, qqq$ and $gg$ bound states. Specific hadrons are identified by
the correspondence of the amplitude in the fifth dimension with the
twist dimension of the interpolating operator for the hadron's
valence Fock state, including its orbital angular momentum
excitations.   An interesting  aspect of our approach is to show
that the mass parameter $\mu R$ which appears in the string theory
in the fifth dimension is quantized, and that it appears as a
Casimir constant governing the orbital angular momentum of the
hadronic constituents analogous to $L(L+1)$ in the radial
Schr\"odinger equation.

\begin{figure}[htb]
\centering
\includegraphics[angle=0,width=10.6cm]{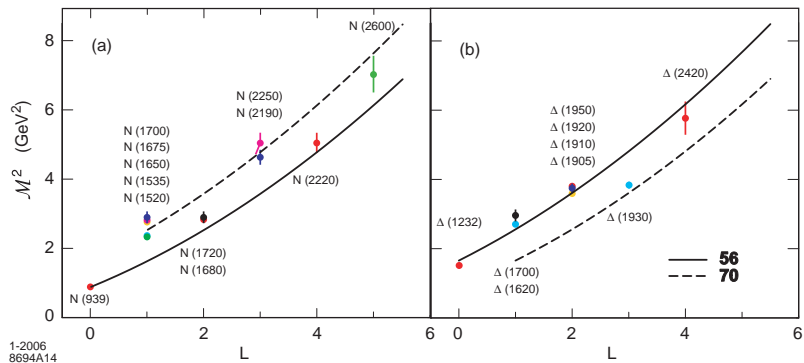}
\caption{Predictions for the light baryon orbital spectrum for
$\Lambda_{QCD}$ = 0.25 GeV. The  $\bf 56$ trajectory corresponds to
$L$ even  $P=+$ states, and the $\bf 70$ to $L$ odd  $P=-$ states.}
\label{fig:BaryonSpec}
\end{figure}

As an example, the set of three-quark baryons with spin 1/2 and
higher  is described  in AdS/CFT by the Dirac equation in the fifth
dimension\cite{Brodsky:2006uq}
\begin{equation}
\left[z^2~ \partial_z^2 - 3 z~ \partial_z
+ z^2 \mathcal{M}^2 - \mathcal{L}_\pm^2 + 4\right] \psi_\pm(z) = 0.
\end{equation}
The constants $\mathcal{L}_+  = L + 1$, $\mathcal{L}_- = L + 2$ in
this equation are  Casimir constants which are determined to match
the twist dimension of the solutions with arbitrary relative orbital
angular momentum. The solution is\begin{equation}
\label{eq:DiracAdS} \Psi(x,z) = C e^{-i P \cdot x} \left[\psi(z)_+~
u_+(P) + \psi(z)_-~ u_-(P) \right],
\end{equation}
with $\psi_+(z) = z^2 J_{1+L}(z \mathcal{M})$ and $\psi_-(z) = z^2
J_{2+L}(z \mathcal{M})$. The physical string solutions have plane
waves and chiral spinors $u(P)_\pm$ along the Poincar\'e coordinates
and hadronic invariant mass states given by $P_\mu P^\mu =
\mathcal{M}^2$. A discrete  four-dimensional spectrum follows when
we impose the boundary condition $\psi_\pm(z=1/\Lambda_{\rm QCD}) =
0$: $\mathcal{M}_{\alpha, k}^+ = \beta_{\alpha,k} \Lambda_{\rm QCD},
~~ \mathcal{M}_{\alpha, k}^- = \beta_{\alpha + 1,k} \Lambda_{\rm
QCD}$, with a scale-independent mass ratio\cite{deTeramond:2005su}.
Figure \ref{fig:BaryonSpec}(a) shows the predicted orbital spectrum
of the nucleon states and Fig.~\ref{fig:BaryonSpec}(b) the $\Delta$
orbital resonances. The spin 3/2 trajectories are determined from
the corresponding Rarita-Schwinger  equation. The data for the
baryon spectra are from S. Eidelman {\em et
al.}\cite{Eidelman:2004wy}   The internal parity of states is
determined from the SU(6) spin-flavor symmetry.
Since only one parameter, the QCD mass scale $\Lambda_{QCD}$, is
introduced, the agreement with the pattern of physical states is
remarkable. In particular, the ratio of $\Delta$ to nucleon
trajectories is determined by the ratio of zeros of Bessel
functions.  The predicted mass spectrum in the truncated space model
is linear $M \propto L$ at high orbital angular momentum, in
contrast to the quadratic dependence $M^2 \propto L$ in the usual
Regge parametrization.

\section{Hadron Wavefunctions in AdS/QCD}

One of the important tools in atomic physics is the Schr\"odinger
wavefunction; it  provides a quantum mechanical description of the
position and spin coordinates of nonrelativistic bound states at a
given time $t$. Similarly, it is  an important goal in hadron and
nuclear physics  to determine the wavefunctions of hadrons in terms
of their fundamental quark and gluon constituents.  The dynamics of
higher Fock states such as the $\vert uud q \bar Q \rangle$
fluctuation of the proton is nontrivial, leading to asymmetric
$s(x)$ and $\bar s(x)$ distributions, $\bar u(x) \ne \bar d(x)$, and
intrinsic heavy quarks $c \bar c$ and $b \bar b$ which have their
support at high momentum\cite{Brodsky:2000sk} . Color adds an extra
element of complexity: for example there are five-different color
singlet combinations of six $3_C$ quark representations which appear
in the deuteron's valence wavefunction, leading to the hidden color
phenomena\cite{Brodsky:1983vf}.

An important example of the utility of light-front wavefunctions in
hadron physics is the computation of polarization effects  such as
the single-spin azimuthal asymmetries  in semi-inclusive deep
inelastic scattering, representing the correlation of the spin of
the proton target and the virtual photon to hadron production plane:
$\vec S_p \cdot \vec q \times \vec p_H$.  Such asymmetries are
time-reversal odd, but they can arise in QCD through phase
differences in different spin amplitudes. In fact, final-state
interactions from gluon exchange between the outgoing quarks and the
target spectator system lead to single-spin asymmetries in
semi-inclusive deep inelastic lepton-proton scattering  which  are
not power-law suppressed at large photon virtuality $Q^2$ at fixed
$x_{bj}$.\cite{Brodsky:2002cx} In contrast to the SSAs arising from
transversity and the Collins fragmentation function, the
fragmentation of the quark into hadrons is not necessary; one
predicts a correlation with the production plane of the quark jet
itself. Physically, the final-state interaction phase arises as the
infrared-finite difference of QCD Coulomb phases for hadron wave
functions with differing orbital angular momentum.  The same proton
matrix element which determines the spin-orbit correlation $\vec S
\cdot \vec L$  also produces the anomalous magnetic moment of the
proton, the Pauli form factor, and the generalized parton
distribution $E$ which is measured in deeply virtual Compton
scattering. Thus the contribution of each quark current to the SSA
is proportional to the contribution $\kappa_{q/p}$ of that quark to
the proton target's anomalous magnetic moment $\kappa_p = \sum_q e_q
\kappa_{q/p}$.\cite{Brodsky:2002cx,Burkardt:2004vm}  The HERMES
collaboration has recently measured the SSA in pion
electroproduction using transverse target
polarization.\cite{Airapetian:2004tw}  The Sivers and Collins
effects can be separated using planar correlations; both
contributions are observed to contribute, with values not in
disagreement with theory
expectations.\cite{Airapetian:2004tw,Avakian:2004qt}

We have recently shown that the  amplitude $\Phi(z)$ describing  the
hadronic state in $\rm{AdS}_5$ can be precisely mapped to the
light-front wavefunctions $\psi_{n/h}$ of hadrons in physical
space-time\cite{Brodsky:2006uq},  thus providing a relativistic
description of hadrons in QCD at the amplitude
level.  The light-front wavefunctions are
relativistic and frame-independent generalizations of the familiar
Schr\"odinger wavefunctions of atomic physics, but they are
determined at fixed light-cone time $\tau= t +z/c$---the ``front
form" advocated by Dirac---rather than at fixed ordinary time $t$.

Formally, the light-front expansion is constructed by quantizing QCD
at fixed light-cone time \cite{Dirac:1949cp} $\tau = t + z/c$ and
forming the invariant light-front Hamiltonian: $ H^{QCD}_{LF} = P^+
P^- - {\vec P_\perp}^2$ where $P^\pm = P^0 \pm
P^z$.\cite{Brodsky:1997de}   The momentum generators $P^+$ and $\vec
P_\perp$ are kinematical; {\em i.e.}, they are independent of the
interactions. The generator $P^- = i {d\over d\tau}$ generates
light-cone time translations, and the eigen-spectrum of the Lorentz
scalar $ H^{QCD}_{LF}$ gives the mass spectrum of the color-singlet
hadron states in QCD together with their respective light-front
wavefunctions.  For example, the proton state satisfies: $
H^{QCD}_{LF} \ket{\psi_p} = M^2_p \ket{\psi_p}$.

\begin{figure}[htb]
\centering
\includegraphics[angle=0,width=10.6cm]{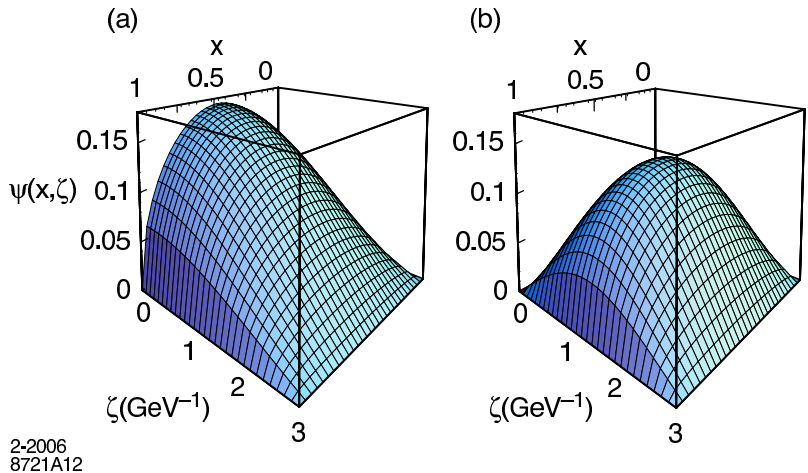}
\caption{AdS/QCD Predictions for the $L=0$ and $L=1$ LFWFs of a meson.}
\label{fig:MesonLFWF}
\end{figure}

\begin{figure}[htb]
\centering
\includegraphics[angle=0,width=10.6cm,height=2.75in]{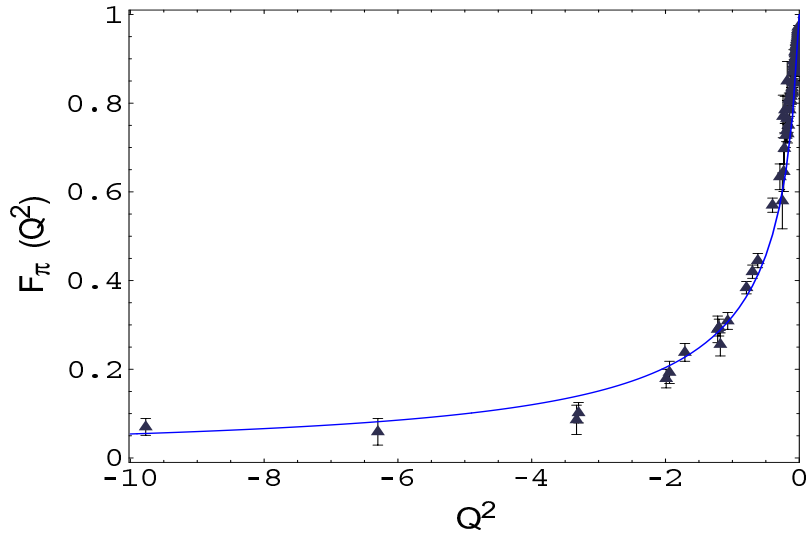}
\caption{AdS/QCD Predictions for the pion form factor.}
\label{fig:PionFF}
\end{figure}

Our approach shows that  there is an exact correspondence between
the fifth-dimensional coordinate of anti-de Sitter space $z$ and a
specific impact variable $\zeta$ in the light-front formalism which
measures the separation of the constituents within the hadron in
ordinary space-time. We derived this correspondence by noticing that
the mapping of $z \to \zeta$ analytically transforms the expression
for the form factors in AdS/CFT to the exact Drell-Yan-West
expression in terms of light-front wavefunctions. In the case of a
two-parton constituent bound state the correspondence between the
string amplitude $\Phi(z)$ and the light-front wave function
$\widetilde\psi(x,\mathbf{b})$ is expressed in closed
form\cite{Brodsky:2006uq}
\begin{equation}  \label{eq:Phipsi}
\left\vert\widetilde\psi(x,\zeta)\right\vert^2 =
\frac{R^3}{2 \pi} ~x(1-x)~ e^{3 A(\zeta)}~
\frac{\left\vert \Phi(\zeta)\right\vert^2}{\zeta^4},
\end{equation}
where $\zeta^2 = x(1-x) \mathbf{b}_\perp^2$. Here $b_\perp$ is the
impact separation and Fourier conjugate to $k_\perp$. The variable
$\zeta$, $0 \le \zeta \le \Lambda^{-1}_{\rm QCD}$, represents the
invariant separation between point-like constituents, and it is also
the holographic variable $z$ in AdS; {\em i.e.}, we can identify
$\zeta = z$. The prediction for the meson light-front wavefunction
is shown in Fig.~\ref{fig:MesonLFWF}. We can also transform the
equation of motion in the fifth dimension using the $z$ to $\zeta$
mapping to obtain an effective two-particle light-front radial
equation
\begin{equation}
\label{eq:Scheq} \left[-\frac{d^2}{d \zeta^2} + V(\zeta) \right]
\phi(\zeta) = \mathcal{M}^2 \phi(\zeta),
\end{equation}
with the effective potential $V(\zeta) \to - (1-4 L^2)/4\zeta^2$ in
the conformal limit. The solution to (\ref{eq:Scheq}) is $\phi(z) =
z^{-\frac{3}{2}} \Phi(z) = C z^\frac{1}{2} J_L(z \mathcal{M}).$ This
equation reproduces the AdS/CFT solutions. The lowest stable state
is determined by the Breitenlohner-Freedman
bound\cite{Breitenlohner:1982jf} and its eigenvalues by the boundary
conditions at $\phi(z = 1/\Lambda_{\rm QCD}) = 0$ and given in terms
of the roots of the Bessel functions: $\mathcal{M}_{L,k} =
\beta_{L,k} \Lambda_{\rm QCD}$. Normalized LFWFs
follow from (\ref{eq:Phipsi})
\begin{equation}
\widetilde \psi_{L,k}(x, \zeta)
=  B_{L,k} \sqrt{x(1-x)}
J_L \left(\zeta \beta_{L,k} \Lambda_{\rm QCD}\right)
\theta\big(z \le \Lambda^{-1}_{\rm QCD}\big),
\end{equation}
where $B_{L,k} = \pi^{-\frac{1}{2}} {\Lambda_{\rm QCD}} \
J_{1+L}(\beta_{L,k})$. The resulting wavefunctions (see:
Fig.~\ref{fig:MesonLFWF}) display confinement at large inter-quark
separation and conformal symmetry at short distances, reproducing
dimensional counting rules for hard exclusive processes and the
scaling and conformal properties of the LFWFs at high relative
momenta in agreement  with perturbative QCD.
The hadron form factors can be predicted from overlap  integrals in
AdS space or equivalently by using the Drell-Yan West formula in
physical space-time.  The prediction for the  pion form factor is
shown in Fig.~\ref{fig:PionFF}.

Since they are complete and orthonormal, these AdS/CFT model
wavefunctions can be used as an initial ansatz for a variational
treatment or as a basis for the diagonalization of the light-front
QCD Hamiltonian.  We are now in fact investigating this possibility
with J. Vary and A. Harindranath. The wavefunctions predicted by
AdS/QCD have many phenomenological applications ranging from
exclusive $B$ and $D$ decays, deeply virtual Compton scattering and
exclusive reactions such as form factors, two-photon processes, and
two body scattering. A connection between the theories and tools
used in string theory and the fundamental constituents of matter,
quarks and gluons, has thus been found.

\bigskip

\noindent{\bf Acknowledgments}

\vspace{5pt} Work supported by the Department of Energy under contract
number DE--AC02--76SF00515 and done in collaboration with Guy de
Teramond.

\end{document}